\documentclass[pra,superscriptaddress,notitlepage,nofootinbib,reprint,reqno]{revtex4-1}
             \usepackage{amssymb,amsfonts,amsmath}
             \usepackage[utf8]{inputenc}
             \usepackage{graphicx}
             \usepackage{mathrsfs}
             \usepackage{braket}
             \usepackage[usenames, dvipsnames,svgnames,table]{xcolor}
             \usepackage[colorlinks=true, allcolors=NavyBlue, breaklinks=true]{hyperref}
             \usepackage{siunitx}
             \usepackage{tikz}
             \usetikzlibrary{decorations.markings,arrows.meta}
             \usepackage{pgfplots}
             \usepgfplotslibrary{external}
\usepackage{dsfont} 

\newcommand{\id}{\mathds{1}}

\DeclareMathOperator{\tr}{tr}

\let\originalleft\left
\let\originalright\right
\renewcommand{\left}{\mathopen{}\mathclose\bgroup\originalleft}
\renewcommand{\right}{\aftergroup\egroup\originalright}
\begin{document}
\title{Quantum superpositions of ``common-cause'' and ``direct-cause'' causal structures}
\author{Adrien Feix}
\author{{\v C}aslav Brukner}
\affiliation{Faculty of Physics, University of Vienna, Boltzmanngasse 5, 1090 Vienna, Austria}
\affiliation{Institute for Quantum Optics and Quantum Information (IQOQI), Boltzmanngasse 3, 1090 Vienna, Austria}
\date{\today}
\begin{abstract}
The constraints arising for a general set of causal relations, both classically and quantumly, are still poorly understood. As a step in exploring this question, we consider a \emph{coherently controlled superposition} of ``direct-cause'' and ``common-cause'' relationships between two events. We propose an implementation involving the spatial superposition of a mass and general relativistic time dilation. Finally, we develop a computationally efficient method to distinguish such genuinely quantum causal structures from classical (incoherent) mixtures of causal structures and show how to design experimental verifications of the nonclassicality of a causal structure.
\end{abstract}
\maketitle
\section{Introduction}\label{sec:introduction}
The deeply rooted intuition that the basic building blocks of the world are \emph{cause-effect-relations} goes back over a thousand years~\cite{aristotle_metaphysics_1933,hume_enquiry_1975,reichenbach_direction_1956} and yet still puzzles philosophers and scientists alike.

In physics, general relativity provides a theoretic account of the causal relations that describe which events in spacetime can influence which other events. For two (infinitesimally close) events separated by a time-like or light-like interval, one event is in the future light cone of the other, such that there could be a direct cause-effect relationship between them. When a space-like interval separates two events, no event can influence the other. The causal relations in general relativity are \emph{dynamical}, since they are imposed by the dynamical light cone structure~\cite{brown_behaviour_2009}.

Incorporating the concept of causal structure in the quantum framework leads to novelties: it is expected that such a notion will be both \emph{dynamical}, as in general relativity, as well as \emph{indefinite}, due to quantum theory~\cite{hardy_quantum_2007}. One might then expect indefiniteness with respect to the question of whether an interval between two events is time-like or space-like, or even whether event $A$ is prior to or after event $B$ for time-like separated events. Yet, finding a unified framework for the two theories is notoriously difficult and the candidate models still need to overcome technical and conceptual problems. 

One possibility to separate conceptual from technical issues is to consider more general, \emph{theory-independent} notions of causality. The \emph{causal model} formalism~\cite{spirtes_causation_1993,pearl_causality:_2000} is such an approach, which has found applications in areas as diverse as medicine,  social sciences and machine learning~\cite{illari_causality_2011}. The study of its quantum extension, allowing for non-local correlations~\cite{wood_lesson_2015,cavalcanti_modifications_2014,henson_theory-independent_2014,fritz_beyond_2016} or including new information-theoretic principles~\cite{pienaar_graphseparation_2015,chaves_informationtheoretic_2015,costa_quantum_2015} might provide intuitions and insights that are currently missing from the theory-laden take at combining quantum mechanics with general relativity.

Recently, it was found that it is possible to formulate quantum mechanics without any reference to a global causal structure~\cite{oreshkov_quantum_2012}. The resulting framework---the \emph{process matrix formalism}---allows for processes which are incompatible with any definite order between operations. One particular case of such a process is the ``quantum switch'', where an auxiliary quantum system can coherently control the order in which operations are applied~\cite{chiribella_quantum_2013}. This results in a quantum controlled superposition of the processes ``$A$ causing $B$'' and ``$B$ causing  $A$''. The quantum switch can also be realized through a preparation of a massive system in a superposition of two distinct states, each yielding a different but definite causal structure for future events~\cite{zych_quantum_2015,zych_bell_????}. Furthermore, it provides computational~\cite{araujo_computational_2014a} and communication~\cite{feix_quantum_2015,allard_guerin_exponential_2016} advantages over standard protocols with a fixed order of events. The first experimental proof-of-principle demonstration of the switch has been reported recently~\cite{procopio_experimental_2015}.

Given that one can implement superpositions of two different causal orders, one may ask if and how one could realize situations in which two events are in superpositions of being in ``common-cause'' ($A$ does not cause $B$ directly) and ``direct-cause'' ($A$ and $B$ share no common cause) relationships. Here we show that such superpositions exist and how to verify them.

We develop a framework for the computationally efficient verification of \emph{coherent superpositions} of ``direct-cause'' and ``common-cause'' causal structures. We propose a natural physical realization of a quantum causal structure with  the spatial superposition of a mass and general relativistic time dilation using the approach developed in Refs.~\cite{zych_bell_????,zych_quantum_2015}. Finally, using the process matrix formalism, we define a degree of ``nonclassicality of causal structures'' and show how to design \emph{experimental verifications} thereof using a semidefinite program~\cite{nesterov_interior_1987}.

\section{Quantum causal models}
\label{sec:causalstructures}
To formalize the pre-theoretic notion of causality, the standard approach is to use \emph{causal models}~\cite{spirtes_causation_1993,pearl_causality:_2000}, consisting of (i) a causal network and (ii) model parameters. The \emph{causal network} is represented by a \emph{directed graph}, whose nodes are variables and whose directed edges represent causal influences between variables. The causal influence from $A$ to $B$ is identified with the possibility of \emph{signaling} from $A$ to $B$. To exclude the possibility of causal loops, one imposes the condition that the graph should be \emph{acyclic} (a ``DAG''), which induces a \emph{partial order} (``causal order'') over the variables. The \emph{model parameters} then determine how the probability distribution of each variable or set of variables is to be computed as a function of the value of its parent nodes.

Fully characterizing the causal model requires information which is available only through ``interventions'', where the value of one or more variables is \emph{set} to take a specific value, independently of the values of the rest of the variables. In the resulting causal network, the connections from all its parents are eliminated. Intervening on all relevant variables is sufficient to completely reconstruct the full causal model~\cite{pearl_causality:_2000}. Since this is often practically impossible, it is crucial to investigate the possibilities of causal inference from a \emph{limited} set of interventions.

Moving to \emph{quantum} causal models, we will define variables as results of generalized quantum operations applied to incoming quantum systems (``local operation''). Formally, a local operation $\mathcal{M}_A: A_I \to A_O$ is a map from a density matrix $\rho_{A_I} \in A_I$ to $\rho_{A_O} \in A_O$ (where $A_I$ ($A_O$) denotes the space of linear operators on the Hilbert space $\mathcal H^{A_I}$ ($\mathcal H^{A_O}$)). The Choi-Jamio\l{}kowski (CJ) isomorphism~\cite{choi_completely_1975,jamiolkowski_linear_1972} provides a convenient representation of the local map as a positive operator $M_A \in A_I \otimes A_O$ (the explicit definition is given in Appendix~\ref{app:cj}).

The \emph{quantum causal structure}, which is the quantum analogue of the classical causal network, maps the aforementioned local operations to a probability distribution. It can be thought of as a \emph{higher order operator} and can be formally represented in the ``superoperator'', ``quantum comb'' or ``process matrix'' formalisms~\cite{gutoski_general_2007,chiribella_quantum_2008,chiribella_theoretical_2009,bisio_quantum_2011,leifer_formulation_2013,oreshkov_quantum_2012}.

\begin{figure}[htp]
  \centering
    \begin{minipage}{.5\linewidth}
    \includegraphics{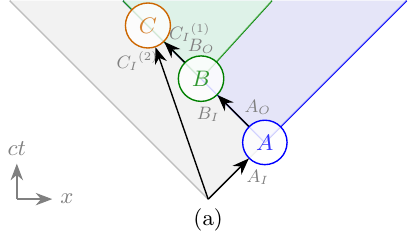}
    \end{minipage}\begin{minipage}{.5\linewidth}
    \includegraphics{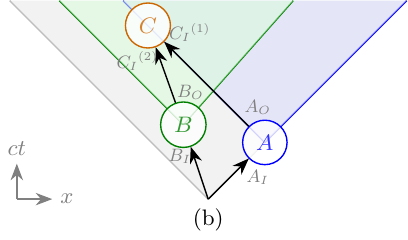}
    \end{minipage}
  \caption{Space-time diagram of two causal structures compatible with the causal order $A\prec B \prec C$: (a) direct-cause process $W^{\text{dc}}$ with a quantum channel between $A_O$ and $B_I$; (b) common-cause process $W^{\text{cc}}$ with a shared (possibly entangled state) between $A_I$ and $B_I$, but no channel between $A_O$ and $B_I$ ($A$ and $B$ are space-like separated).}
  \label{fig:a-b-c-causal networks}
\end{figure}

We will focus on quantum causal structures with three laboratories (three nodes in the graph) $A$, $B$ and $C$ compatible with the causal order ``A is not after B, which is not after C'' ($A \prec B \prec C$). This means that there are no causal influences from $B$ and $C$ to $A$, nor from $C$ to $B$ (see Fig.~\ref{fig:a-b-c-causal networks}). (Since $C$ is last, $C$'s output space $C_O$ can be disregarded.)

In the process matrix formalism, the quantum causal structure is represented by the matrix $W \in A_I \otimes A_O \otimes B_I \otimes B_O \otimes C_I$~\cite{oreshkov_quantum_2012,araujo_witnessing_2015}. The probabilities of observing the outcomes $i,j,k$ at $A, B, C$ (corresponding to implementing the completely positive (CP) maps $M_A^{i}$, $M_B^{j}$, $M_C^{k}$ respectively) are given by the \emph{generalized Born rule}:
\begin{equation}\label{eq:gen-born-rule}
p(A=i,B=j,C=k) = \tr[W\, (M_A^{i} \otimes M_B^{j} \otimes M_C^{k})].
\end{equation}

The quantum causal structure and local operations should generate only meaningful (that is, \emph{positive} and \emph{normalized}) probability distributions. In addition, we require the probability distributions to be compatible with the causal order $A \prec B \prec C$. Note that both ``common-cause'' and ``direct-cause'' relationships between $A$ and $B$ are compatible with this causal order.

In terms of process matrices, these conditions are equivalent to requiring that $W$ satisfies~\cite{araujo_witnessing_2015}:
\begin{gather}\label{eq:process_representation}
W \geq 0, \quad W = \mathcal L_{A\prec B\prec C} (W)\\
\tr W = d_{A_O} d_{B_O} \label{eq:process_representation4}.
\end{gather}
$\mathcal L_{A\prec B\prec C} (\cdot)$ is the projection onto processes compatible with the causal order $A \prec B \prec C$, defined in Appendix~\ref{app:causallyordered}. Eq.~\eqref{eq:process_representation} defines a convex cone $\mathcal W$, eq.~\eqref{eq:process_representation4} a normalization constraint. 

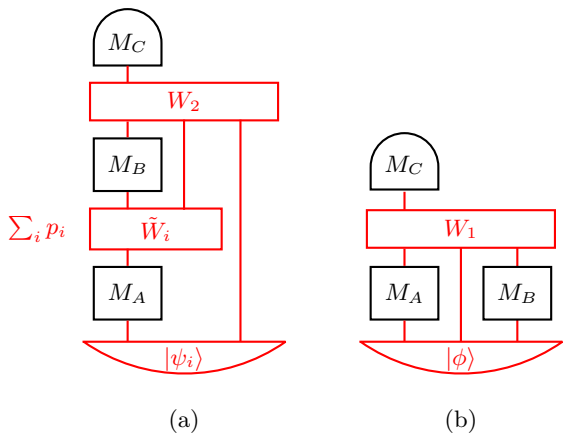
\begin{figure}[htp]
    \centering
      \begin{tikzpicture}[scale=1.5]
                \draw[thick, red] (0.9,-0.7) arc (125:55:-1.56) (-0.9,-0.7) -- (0.9,-0.7);
                \node[red] at (0,-0.85) {$\ket{\psi_i}$};
                \draw[thick, red]  (0,0.47) -- (0,1.27);
                \draw[thick, red]  (0.5,-0.7) -- (0.5,1.27);
        \node[draw, thick, rectangle,minimum width=0.9cm,minimum height=0.7cm, fill=none] (A) at (-0.5,-0.27) {${M}_{A}$};
        \node[draw, thick, rectangle,minimum width=0.9cm,minimum height=0.7cm, fill=none] (B) at (-0.5,0.87) {${M}_{B}$};
        \draw [thick] (-0.8,1.95) -- (-0.8,1.75) -- (-0.2,1.75) -- ++(0,0.2) arc (0:180:0.3);
                \node (C) at (-0.5,1.95) {$M_C$}; 
        \node[draw, thick, red, rectangle,minimum width=1.75cm,minimum height=0.5cm, fill=none] (U) at (-0.25,0.3) {$\tilde{W}_i$};
        \node[draw, thick, red, rectangle,minimum width=2.5cm,minimum height=0.5cm, fill=none] (U2) at (0,1.43) {$W_{2}$};
        \draw[thick, red] (-0.5,-0.7) -- ++(0,0.188);
        \draw[thick, red] (A) -- ++(0,0.395); \draw [thick,red] (B) -- ++(0,0.383);
        \draw [thick, red] (B) -- ++(0,-0.395); \draw [thick,red] (-0.5,1.748) -- ++(0,-0.15);
        \node[red] (sum) at (-1.3,0.3) {$\sum_i p_i$};
                \node[] (label) at (0,-1.4) {(a)};
  \end{tikzpicture}
        \begin{tikzpicture}[scale=1.5]
                \draw[thick, red] (0.9,-0.7) arc (125:55:-1.56) (-0.9,-0.7) -- (0.9,-0.7);
                \node[red] at (0,-0.85) {$\ket{\phi}$};
                \draw[thick, red] (0.5,-0.7) -- ++(0,0.84) (0,-0.7) -- ++(0,0.84);
        \node[draw, thick, rectangle,minimum width=0.9cm,minimum height=0.7cm, fill=none] (A) at (-0.5,-0.27) {${M}_{A}$};
        \node[draw, thick, rectangle,minimum width=0.9cm,minimum height=0.7cm, fill=white] (B) at (0.5,-0.27) {${M}_{B}$};
        \draw [thick] (-0.8,0.85) -- (-0.8,0.65) -- (-0.2,0.65) -- ++(0,0.2) arc (0:180:0.3);
                \node (C) at (-0.5,0.87) {$M_C$}; 
        \node[draw, thick, red, rectangle,minimum width=2.5cm,minimum height=0.5cm, fill=none] (U) at (0,0.3) {$W_{1}$};
        \draw[thick, red] (-0.5,-0.7) -- ++(0,0.188);
        \draw[thick, red] (A) -- ++(0,0.395); \draw [thick,red] (-0.5,0.47) -- ++(0,0.16);
                \node[] (label) at (0,-1.4) {(b)};
                \node[] (hidden) at (-1.4,0) {};
                \node[] (hidden) at (1.1,0) {};
  \end{tikzpicture}
    \caption{Circuit representation of the causal structures of Fig.~\ref{fig:a-b-c-causal networks}, where $\ket{\psi_i}$ and $\ket{\phi}$ are states, $\tilde{W_i}, W_2$ and $W_1$ are CP trace preserving (CPTP) maps (lines can represent quantum systems of different dimensions). (a) The direct-cause process $W^\text{dc}$ is the most general one satisfying \eqref{eq:w-direct-cause}; (b) the common-cause process $W^\text{cc}$ is the most general one satisfying \eqref{eq:w-common-cause}.}
    \label{fig:a-b-c-circuits}
\end{figure}

Following the standard DAG terminology, a purely ``direct-cause'' process $W^\text{dc}$ contains only a \emph{direct cause-effect relation between A and B}, excluding any form of \emph{common cause} between $A$ and $B$. Any correlation between $A$ and $B$ is therefore caused by $A$ alone (Fig.~\ref{fig:a-b-c-causal networks}\,(a) and Fig.~\ref{fig:a-b-c-circuits}\,(a)). Tracing out $C_I$ and $B_O$, the process matrix is a tensor product $\rho^{A_{I}} \otimes \tilde {W}^{A_{O} B_{I}}$. In our scenario, it will prove natural to \emph{extend} this definition to include \emph{convex mixtures} of direct-cause processes, i.e.,
\begin{equation}
\label{eq:w-direct-cause}
\tr_{C_{I} B_O} W^{\text{dc}} = \sum_i p_i \rho_i^{A_{I}} \otimes \tilde {W}_i^{A_{O} B_{I}},
\end{equation}
where $p_i \geq 0, \sum_i p_i =1$, $\rho^{A_{I}}_i$ are arbitrary states and $\tilde{W}_i^{A_{O} B_{I}}$ arbitrary valid channels between Alice's output and Bob's input, representing to direct cause-effect links between $A$ and $B$. 

Such a process can be interpreted as a probability distribution over states entering $A_I$ and corresponding channels from $A_O$ to $B_I$. In the DAG framework, such probability distributions can be obtained from a graph with an additional latent node that acts as a common cause for all the observed nodes or simply ignorance of the graph that is implemented. Every channel from $A$ to $B$ with classical memory can be decomposed in this way; see Appendix~\ref{app:def-cause-effect} for details.

On the other hand, a purely ``common-cause'' process $W^{\text{cc}}$ does not include \emph{any direct causal influence between A and B} (Fig.~\ref{fig:a-b-c-causal networks}\,(b) and Fig.~\ref{fig:a-b-c-circuits}\,(b)). This implies that there is no channel between $A_O$ and $B_I$. Therefore, when $B_O$ and $C_I$ are traced out, the process factorizes as
\begin{equation}
\label{eq:w-common-cause}
\tr_{C_{I} B_O} W^{\text{cc}} = \sigma^{A_{I} B_I} \otimes \id^{A_O},
\end{equation}
where $\sigma^{A_{I} B_I}$ is an arbitrary (possibly entangled, possibly mixed) state, representing the common-cause influencing $A$ and $B$.

\section{Classical and quantum superpositions of causal structures}
\label{sec:coherent-sup}
One possibility of combining direct-cause and common-cause processes consists in allowing for \emph{classical mixtures} thereof: imagine that flipping a (possibly biased) coin determines which process will be realized in an experimental run. Formally, this is described by a process $W^\text{conv}$ which can be decomposed as a convex combination:
\begin{equation}
\label{eq:w-separable}
W^{\text{conv}} = q W^{\text{cc}} + (1-q) W^{\text{dc}},
\end{equation}
where $0 \leq q \leq 1$, $W^{\text{dc}}$ satisfies~\eqref{eq:w-direct-cause} and $W^{\text{cc}}$ satisfies~\eqref{eq:w-common-cause}.  Note that such a classical mixture was experimentally implemented in Ref.~\cite{ried_quantum_2015}.

Can there be causal structures exhibiting \emph{genuine quantum coherence}, i.e., that cannot be decomposed as a classical mixture of direct-cause and common-cause processes (while respecting the causal order $A\prec B \prec C$)?

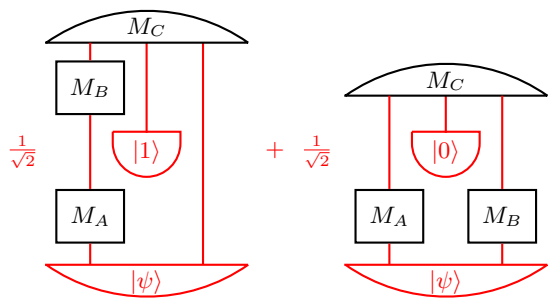
\begin{figure}[htp]
    \centering
      \begin{tikzpicture}[scale=1.5]
                \draw[thick, red] (-0.3,0.48) -- ++(0,-0.1) arc (-180:0:0.3) -- ++(0,0.1) -- ++ (-0.6,0);
                \node[red] at (0,0.3) {$\ket{1}$};
                \draw[thick, red] (0.9,-0.7) arc (125:55:-1.56) (-0.9,-0.7) -- (0.9,-0.7);
                \node[red] at (0,-0.85) {$\ket{\psi}$};
                \draw[thick, red]  (0,0.47) -- (0,1.27);
                \draw[thick, red]  (0.5,-0.7) -- (0.5,1.27);
        \node[draw, thick, rectangle,minimum width=0.9cm,minimum height=0.7cm, fill=none] (A) at (-0.5,-0.27) {${M}_{A}$};
        \node[draw, thick, rectangle,minimum width=0.9cm,minimum height=0.7cm, fill=none] (B) at (-0.5,0.87) {${M}_{B}$};
        \draw[thick] (0.9,1.27) arc (55:125:1.56) (-0.9,1.27) -- (0.9,1.27);
                \node (C) at (0,1.41) {$M_C$}; 
        \draw[thick, red] (-0.5,-0.7) -- ++(0,0.188);
        \draw[thick, red] (A) --  (B) ;
        \draw [thick, red] (B) -- (-0.5,1.27);
               \node[red] (sum) at (-1.1,0.3) {$\frac{1}{\sqrt{2}}$};
  \end{tikzpicture}
        \begin{tikzpicture}[scale=1.5]
                \draw[thick, red] (-0.3,0.48) -- ++(0,-0.1) arc (-180:0:0.3) -- ++(0,0.1) -- ++ (-0.6,0);
                \node[red] at (0,0.3) {$\ket{0}$};
                \draw[thick, red] (0.9,-0.7) arc (125:55:-1.56) (-0.9,-0.7) -- (0.9,-0.7);
                \node[red] at (0,-0.85) {$\ket{\psi}$};
                \draw[thick, red]  (0,0.47) -- (0,0.8);
        \node[draw, thick, rectangle,minimum width=0.9cm,minimum height=0.7cm, fill=none] (A) at (-0.5,-0.27) {${M}_{A}$};
        \node[draw, thick, rectangle,minimum width=0.9cm,minimum height=0.7cm, fill=none] (B) at (0.5,-0.27) {${M}_{B}$};
        \draw[thick] (0.9,0.8) arc (55:125:1.56) (-0.9,0.8) -- (0.9,0.8);
                \node (C) at (0,0.93) {$M_C$}; 
        \draw[thick, red] (-0.5,-0.7) -- ++(0,0.188);
        \draw[thick, red]  (0.5,-0.7) -- (B);
        \draw [thick, red] (B) -- (0.5,0.8) (A) -- (-0.5,0.8);
        \node[red] (sum) at (-1.3,0.3) {$+~~\frac{1}{\sqrt{2}}$};
                \node[] (hidden) at (0.7,0) {};
  \end{tikzpicture}
  \caption{Coherent superposition of a direct-cause and a common-cause process, implementing the causal structure $W^\text{coherent}$ of \eqref{eq:coherent-pure}.}\label{fig:superposition-circuits}
\end{figure}

We now give an example of such a coherent superposition. It is analogous to the ``quantum switch''~\cite{chiribella_quantum_2013}, which coherently superposes two causal orders $A \prec B \prec C$ and $B \prec A \prec C$, where the causal structure is entangled to a ``control'' system $C_I^{(0)}$ added to $C$'s input space\footnote{See Ref.~\cite{maclean_quantum-coherent_2016} for a different type of quantum causal structure proposed independently.}. To keep the notation simple, we define it in the ``pure'' CJ-vector notation (see Appendix~\ref{app:cj}):
\begin{multline}
\label{eq:coherent-pure}
\ket{w} = \frac{1}{\sqrt{2}}\bigg(\ket{0}^{C_I^{(0)}} \ket{\psi}^{A_{I} B_{I}} \ket{I}\rangle^{A_{O} C_I^{(1)}} \ket{I}\rangle^{B_{O} C_I^{(2)}}  \\
 + \ket{1}^{C_I^{(0)}} \ket{\psi}^{A_{I} C_I^{(2)}} \ket{I}\rangle^{A_{O} B_{I}} \ket{I}\rangle^{B_{O} C_I^{(1)}}\bigg), \\
W^{\text{coherent}} := \ket{w}\bra{w} \quad\quad
\end{multline}
where $\ket{I}\rangle := \sum_{j=1}^{d} \ket{jj}$ represents a non-normalized maximally entangled state---the CJ-representation of an identity channel. The corresponding superposition of circuits is shown in Fig.~\ref{fig:superposition-circuits}. $W^\text{coherent}$ satisfies neither the direct-cause condition \eqref{eq:w-direct-cause} nor the common-cause condition \eqref{eq:w-common-cause} and is a projector on a pure vector, so it cannot be decomposed into \emph{any nontrivial convex combination}, in particular not a mixture of direct-cause and common-cause processes. This proves that the process's causal structure is nonclassical.

\section{Physical implementation of the quantum causal structure}
The causal structure $W^\text{coherent}$ would not be of particular interest if it were a mere theoretical artifact. We now give an explicit and plausible physical scenario to realize the quantum causal structures in models which respect the principles of general relativistic time dilation and quantum superposition. We utilize the approach recently developed for the ``gravitational quantum switch'' to realize a superposition and entanglement of two different causal orders~\cite{zych_quantum_2015,zych_bell_????}.

\begin{figure}[htp]
  \centering
    \begin{minipage}{.5\linewidth}
    \includegraphics{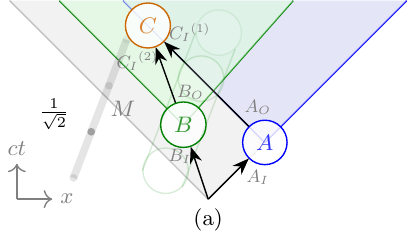}
    \end{minipage}\begin{minipage}{.5\linewidth}
    \includegraphics{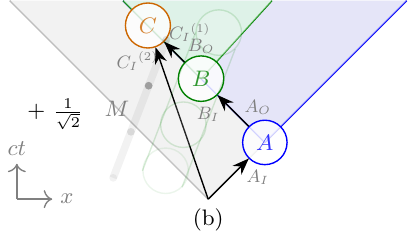}
    \end{minipage}
    \caption{Space-time diagrams of events in a superposition of casual structures, as seen from a distant observer. Bob's laboratory is moving along a time-like curve, indicated by the circles showing his laboratory before and after $\tau_B$. (a) If the mass $M$ is far away from Bob, the event at his local time $\tau_B$ is space-like separated from $A$ and a common-cause causal structure is realized. (b) If $M$ is sufficiently close to $B$, because of time dilation, $B$'s event at time $\tau_B$, is in the future light cone of $A$, establishing a direct-cause structure between $A$, $B$ and $C$. For a \emph{coherent superposition} of the positions of $M$ (the position of $M$ being the control system $C_I^{(0)}$), the quantum causal structure will be described by $W_\text{coherent}$, as given in \eqref{eq:coherent-pure}.}\label{fig:gravitational}
\end{figure}

Consider two observers, Alice and Bob, who have initially synchronized clocks. We \emph{define} the events in the respective laboratories with respect to the \emph{local clocks}. Bob's local operation will always be applied at his local time $\tau_B$, while Alice's is applied at her local time $\tau_A$. We will consider two configurations, which will be controlled by a quantum system. The state of the control system is given by the position of a massive body. In the first configuration, all masses are sufficiently far away such that the parties are in an approximately flat spacetime. The events in the two laboratories are chosen such that the event $B$ is outside of $A$'s light cone and the common-cause causal relationship is implemented. The coordinate times of the two events, as measured by a local clock of a distant observer, are $t_A \approx \tau_A$ and $t_B \approx \tau_B$. (Fig.~\ref{fig:gravitational}\,(a)). 

In the second configuration, a mass $M$ is put closer to Bob's laboratory than to Alice's such that his clock runs slower with respect to hers due to gravitational time dilation. With a suitable choice of mass and distance between Alice and Bob, the event $B$, which is defined by his clock showing local time $\tau_B$, will be inside $A$'s future light cone. In terms of coordinate times one now has $t'_A= \tau_A/ \sqrt{-g_{00}(A)}$ and $t'_B= \tau_B / \sqrt{-g_{00}(B)}$, where $g_{00}(A)$ and $g_{00}(B)$ are the ``00'' components of the metric tensor at the position of the laboratories. This configuration can implement the direct-cause relationship (Fig.~\ref{fig:gravitational}\,(b)). 

If the mass $M$ is initially in a coherent \emph{spatial superposition} of a position close and a position far away from Bob, the quantum superposition of causal structures $W_\text{coherent}$ is implemented. The \emph{position of the mass} acts as the control system $C_I^{(0)}$;\footnote{The state $\ket{0}$ corresponding to the mass being far away from Bob and the state $\ket{1}$ corresponding to the mass being close to Bob.} it can be received by Charlie, who can manipulate it further (in particular, measure it in the superposition basis). Any possible information about the causal structure (direct cause or common cause) encoded in the degrees of freedom of the laboratories, such as for example in the clocks of the labs, must be erased, possibly using the methods of Ref~\cite{zych_bell_????}.

Note that, in contrast to the superposition of different causal orders~\cite{zych_quantum_2015,zych_bell_????}, the time dilation necessary to ``move $B$ in or out'' of the light cone can, in principle, be made \emph{arbitrarily small}, if Bob can define $\tau_B$ and thus the event $B$ with a sufficiently precise clock\footnote{If Bob's clock cannot resolve the interval $\tau_B (1-1/ \sqrt{-g_{00}(B)})$ within the time $\tau_B$, the event $B$ will be inside or outside $A$'s light cone randomly and independently of the position of $M$, adding noise to the process.}. 

To give an idea of the orders of magnitude involved: for a spatial superposition of the order of $\Delta x=\SI{1}{mm}$ and a mass of $M=\SI{1}{g}$, Bob's clock should resolve one part in \SI{e27}{} to be able to certify the nonclassicality of the causal structure. This regime is still quite far from experimental implementation, since the best molecule interferometers~\cite{eibenberger_matterwave_2013} do not go beyond $M=\SI{e5}{amu}$, $\Delta x = \SI{e-6}{m}$, while the best atomic lattice clocks achieve a precision of one part in \SI{e18}~\cite{nicholson_systematic_2015}. An additional difficulty consists in avoiding significant entanglement between the position of the mass and systems other than the local clocks. Nonetheless this regime is still far away from the Planck scale that is usually assumed to be relevant for quantum gravity effects.

We also stress that the process $W_\text{coherent}$, although it cannot be decomposed as a convex combination of a common cause and a direct cause process, is still compatible with the causal order $A \prec B \prec C$ and, as such~\cite{bisio_minimal_2011}, can be realized as a quantum circuit, as shown in Fig.~\ref{fig:w3} (b) of Appendix~\ref{app:causallyordered}. 

\section{Verifying the nonclassicality of causal structures}
\label{sec:witnessing}
We now provide an \emph{experimentally accessible} and \emph{efficiently computable} measure of the nonclassicality of causality.

Let us first define the set $\mathcal S$ of operators which are positive on any convex combination $W^\text{conv}$ of direct-cause and common-cause processes (i.e., processes satisfying \eqref{eq:w-separable}):
\begin{equation}\label{eq:causal-witness}
S \in \mathcal S \Rightarrow \tr [S\, W^{\text{conv}}] \geq 0 \quad \forall W^\text{conv}.
\end{equation}
If $S$ is positive on all convex combinations of direct-cause and common-cause process matrices, then it is also positive on all direct-cause ($\tr [S\, W^{\text{dc}}] \geq 0$) and common-cause ($\tr [S\, W^{\text{cc}}] \geq 0$) processes individually.

Since $W^{\text{dc}}$ is a direct-cause process~\eqref{eq:w-direct-cause} if and only if the operator $\tr_{C_I B_O} W^{\text{dc}}$ is separable with respect to the bipartition $(A_I,A_O B_I)$, we effectively require $S$ to be an \emph{entanglement witness}~\cite{horodecki_quantum_2009,chruscinski_entanglement_2014} of the reduced process for the bipartition $(A_I,A_O B_I)$. The full characterization of the set of entanglement witnesses is known to be computationally hard~\cite{gurvits_classical_2003}. Instead, we will use the \emph{positive partial transpose}~\cite{peres_separability_1996,horodecki_separability_1996} criterion as a relaxation to define an efficiently computable measure of nonclassicality.

Enforcing that $S$ is positive on common-cause process matrices in terms of semidefinite constraints is straightforward: since the condition for $W^{\text{cc}}$ \eqref{eq:w-common-cause} to be a common-cause process is already a semidefinite constraint, the ``dual'' constraint for $S$ to be positive on all common-cause process matrices is semidefinite as well.

The operators in the set $\mathcal S_\text{SDP}$ (explicitly constructed in Appendix~\ref{app:s1s2}) are defined as those that obey \emph{both} the condition of having a positive partial transpose and being positive on all common-cause process matrices. Every $S \in \mathcal S_\text{SDP}$ has positive trace with any $W^\text{conv}$. Conversely, $\tr [S\, W] < 0$ certifies that the process $W$ is a genuinely nonclassical causal structure---the operators $S$ can therefore be used as \emph{nonclassicality of causality witnesses}\footnote{The ``causal witnesses'' introduced in Ref.~\cite{araujo_witnessing_2015} are conceptually different, since they examine whether a process can be decomposed as a convex mixture of \emph{causally ordered} processes. All of the processes we study here have a fixed causal order $A \prec B \prec C$.}.

It is crucial to realize that for every given genuinely quantum $W$, one can \emph{efficiently optimize}---the optimization is a semidefinite program~\cite{nesterov_interior_1987}---over the set of nonclassicality witnesses to find the one that has minimal trace with $W$:
\begin{equation}
\label{eq:sdp-witness-explicit}
\begin{gathered}
\min \tr [S\,W]\\
\text{s.t.} \quad S \in \mathcal S_\text{SDP}, \quad \id/d_{O} - S \in \mathcal W^{ *},
\end{gathered}
\end{equation}
where $\mathcal W^*$ is the dual cone of $\mathcal W$, given in Appendix~\ref{app:dualcone}. The \emph{normalization condition} $\id/d_O - S \in \mathcal W^*$ is necessary for the optimization to reach a finite minimum and confers an operational meaning to $\mathcal C(W) := -\tr[S_\text{opt}\,W]$: it is the amount of ``worst-case noise'' the process can tolerate before its quantum features stop being detectable by witnesses in $\mathcal S_\text{SDP}$ (in analogy to the ``generalized robustness of entanglement''~\cite{steiner_generalized_2003}). Because of its ability to certify the quantum nonclassicality of causal structures, we will refer to $\mathcal C(\cdot)$ as the ``nonclassicality of causality''. Note that $\mathcal C(\cdot)$ satisfies the natural properties of \emph{convexity} and \emph{monotonicity under local operations} (see Appendix~\ref{sec:measure}).

To experimentally verify the properties of a process like $W^\text{coherent}$, one can use the semidefinite program \eqref{eq:sdp-witness-explicit} to compute the optimal nonclassicality of causality witness $S_\text{opt}$ for $W^\text{coherent}$. The nonclassicality of causality $\mathcal C(W^\text{coherent})$ can be measured by decomposing $S_\text{opt}$ in a convenient basis of local operations. In general, this is as demanding as performing a full ``causal tomography''~\cite{araujo_witnessing_2015,ried_quantum_2015,costa_quantum_2015}.

\section{Causal inference under experimental constraints}
\label{sec:exp-constraints}
There are two reasons to consider witnesses that are subject to certain \emph{additional restrictions}. First, there might be various technical limitations arising from the experimental setup~\cite{ried_quantum_2015,procopio_experimental_2015}, which make full tomography impractical. Second, in analogy to the classical case, it is of \emph{conceptual} interest to investigate the power of \emph{quantum causal inference mechanisms} working on \emph{limited data}. In particular, one might want to investigate differences between quantum and classical causal inference algorithms under such constraints.

As an application of this method, we will examine witnesses for the process $W^\text{coherent}$. In the following, we will consider qubit input and output spaces, i.e., $\dim A_{I} = \dim A_{O} = \dim B_{I} = \dim C_{I}^{(0,1,2)} = 2$ for simplicity and computational speed. The optimal witness for $W_{\text{coherent}}$, obtained from the optimization \eqref{eq:sdp-witness-explicit} using YALMIP~\cite{yalmip} with the solver MOSEK~\cite{mosek}, leads to a nonclassicality of causality of $\mathcal C (W^\text{coherent}) = - \tr [S_{\text{opt}} W_{\text{coherent}}] \approx 0.2278$. 

An intriguing feature of quantum causal models is that direct-cause correlations (Fig.~\ref{fig:a-b-c-causal networks}\,(a)) and common-cause correlations (Fig.~\ref{fig:a-b-c-causal networks}\,(b)) can be distinguished through a restricted class of informationally symmetric operations~\cite{leifer_formulation_2013}, sometimes called ``observations''~\cite{fitzsimons_quantum_2013,ried_quantum_2015} that are non-demolition measurements (we refer the reader to Appendix~\ref{app:observations} for certain issues with this definition). We can constrain a witness $S^\text{ndmeas}$ to consist of linear combinations of such non-demolition measurements through an additional condition to the semidefinite program~\eqref{eq:sdp-witness-explicit}, given in Appendix~\ref{app:exp-constraints}.

Surprisingly, purely ``observational'' witnesses are sufficient not only to distinguish common-cause from direct-cause processes, but \emph{also} to distinguish a classical mixture of direct-cause and common-cause processes from a genuine quantum superposition, since $- \tr [S^{\text{ndmeas}}_{\text{opt}} W_{\text{coherent}}] \approx 0.0732$.

Since measurements and repreparations and even non-demolition measurements are often challenging to implement~\cite{grangier_quantum_1998}, it can also be useful to consider a nonclassicality of causality witness $S^\text{unitary}$ which can be decomposed into \emph{unitary operations} for $A$ and $B$, and arbitrary measurements for $C$. The requirement can also easily be translated in a semidefinite constraint, given in Appendix~\ref{app:exp-constraints}. One finds that $- \tr [S^{\text{unitary}}_{\text{opt}} W_{\text{coherent}}] \approx 0.1686$. A summary of the different constraints mentioned in this section can be found in Appendix~\ref{app:exp-constraints}.

\section{Conclusions}
We presented a three-event quantum causal model compatible with the causal order $A \prec B \prec C$ which is a quantum controlled \emph{coherent superposition between common-cause and direct-cause models}, not a classical mixture thereof.

The experimental implementation we proposed is of conceptual interest, since it relies both on general relativity and the quantum superpositions principle, two elements we expect to feature in a full theory unifying quantum theory and general relativity. Interestingly, both the mass of the object and the separation between the two amplitudes can be arbitrarily small, as long as Bob has access to a sufficiently precise clock to define the instant of his event $B$.

In order to experimentally certify a genuinely quantum causal structure, we introduced and characterized \emph{nonclassicality of causality witnesses} and provided a semidefinite program to efficiently compute them. Experimental and conceptual constraints are readily included in the framework.

The potential of quantum causal structures as a quantum information resource was recently demonstrated in terms of query complexity~\cite{araujo_computational_2014a} and communication complexity~\cite{feix_quantum_2015,allard_guerin_exponential_2016}, but is still poorly understood. It would be interesting to understand which advantages could be obtained from the coherent superpositions of and common- and direct-cause processes.

\textit{Remark.---} In the final stages of completing this manuscript, a related work by MacLean et al.~\cite{maclean_quantum-coherent_2016} appeared independently. The difference in the definitions of direct-cause processes between the two papers and its implications are discussed in Appendix~\ref{app:def-cause-effect}.

\textit{Acknowledgements.---} We thank Mateus Araújo, Fabio Costa, Flaminia Giacomini, Nikola Paunkovi\'c, Jacques Pienaar and Katia Ried for useful discussions. We acknowledge support from the Austrian Science Fund (FWF) through the Special Research Programme FoQuS, the Doctoral Programme CoQuS, the project I-2526 and the research platform TURIS. This publication was made possible through the support of a grant from the John Templeton Foundation. The opinions expressed in this publication are those of the authors and do not necessarily reflect the views of the John Templeton Foundation.

\appendix

\section{Choi-Jamio\l{}kowski isomorphism}
\label{app:cj}
The Choi-Jamio\l{}kowski (CJ) representation of a CP map $\mathcal{M}_A: A_I \to A_O$ is
\begin{equation}
\label{eq:cj}
M_A := \left[\left({\cal I}\otimes{\cal M}_{A} \right)(\ket{I}\rangle\langle\bra{I})\right]^{\mathrm T} \in A_I \otimes A_O,
\end{equation}
where $\mathcal I$ is the identity map, $\ket{I}\rangle:= \sum_{j=1}^{d_{\mathcal{H}_{I}}} \ket{jj} \in \mathcal{H}_{I}\otimes \mathcal{H}_{I}$ is a non-normalized maximally entangled state and $^\text{T}$ denotes matrix transposition in the computational basis. 

The inverse transformation is then defined as:
\begin{equation}
\label{eq:cj-inverse}
\mathcal M_A(\rho)= \tr_{I} \left[(\rho \otimes \id) M_A \right]^{\mathrm T}.
\end{equation}

For operations which have a single Kraus operator ($\mathcal M_A(\rho) = A\rho A^\dagger$), one also define a ``pure CJ-isomorphism''~\cite{royer_wigner_1991,braunstein_universal_2000}, which maps the operation to a \emph{vector}\footnote{Note that there are differing conventions, where the conjugation is omitted.}:
\begin{equation}
\label{eq:pure-cj}
\ket{A^*}\rangle := (\id \otimes A^*)\ket{I}\rangle \in \mathcal{H}^{A_I} \otimes \mathcal{H}^{A_O} 
\end{equation}
The usual CJ-representation of such an operation is simply the \emph{projector onto the CJ-vector}: $M_A = \ket{A^*}\rangle\langle\bra{A^*}$.
\section{Causally ordered and common-cause process matrices}
\label{app:causallyordered}
We first introduce a shorthand that we will use throughout the following appendices:
\begin{equation}
\label{def:notation_trace}
 _X W := \frac{\id^{X}}{d_X} \otimes \tr_X W,
\end{equation}
where $d_X$ is the dimension of the Hilbert space $X$.

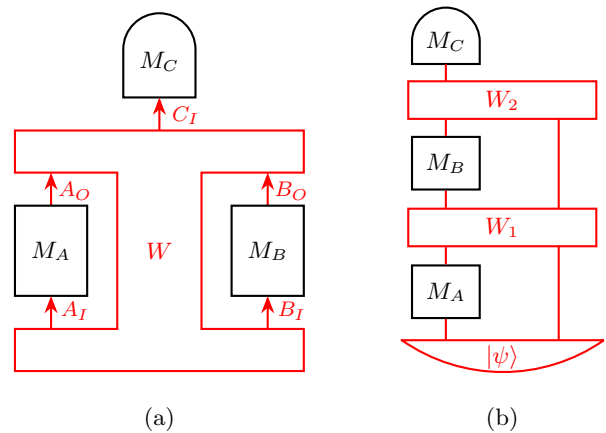
\begin{figure}[htp]
  \centering
         \begin{tikzpicture}[scale=1.6]
        \node[] (alice) at (-0.9,0)  {$M_A$};
        \node[red] (W) at (0,0) {$W$};
        \node[] (bob) at (0.9,0) {$M_B$};
                \node[] (F) at (0,1.275) {};
                 \draw[red, thick] (-1.2,-1) -- (1.2,-1) -- (1.2,-0.65) -- (0.35,-0.65) -- (0.35,0.65) -- (1.2,0.65) -- (1.2,1) -- (-1.2,1) -- (-1.2, 0.65) -- (-0.35,0.65) -- (-0.35,-0.65) -- (-1.2,-0.65) -- cycle;
                \draw[thick, red, -Stealth] (-0.9,0.375) -- (-0.9,0.65);
        \draw[thick] (-1.2,0.375) -- (-0.6,0.375) -- (-0.6,-0.375) -- (-1.2,-0.375) -- cycle;
                \draw[thick, red, Stealth-] (-0.9,-0.375) -- (-0.9,-0.65);
                \draw[thick, red, -Stealth] (0.9,0.375) -- (0.9,0.65);
                \draw[thick] (1.2,0.375) -- (0.6,0.375) -- (0.6,-0.375) -- (1.2,-0.375) -- cycle;
                \draw [thick] (-0.3,1.675) -- (-0.3,1.275) -- (0.3,1.275) -- ++(0,0.4) arc (0:180:0.3);
                \node (charlie) at (0,1.575) {$M_C$};             
                \draw[thick, red, Stealth-] (0.9,-0.375) -- (0.9,-0.65);
                \draw[thick, red, Stealth-] (0,1.275) -- (0,1);
        \node[red] (aliceO) at ([shift={(0.2cm,0.5cm)}]alice) {\footnotesize $A_O$};
        \node[red] (aliceI) at ([shift={(0.2cm,-0.5cm)}]alice) {\footnotesize $A_I$};
        \node[red] (bobO) at ([shift={(0.2cm,0.5cm)}]bob) {\footnotesize $B_O$};
        \node[red] (bobI) at ([shift={(0.2cm,-0.5cm)}]bob) {\footnotesize $B_I$};
        \node[red] (CI) at ([shift={(0.22cm,-0.15cm)}]F) {\footnotesize $C_I$};
        \node[] (label) at (0,-1.4) {(a)};
  \end{tikzpicture}
  \begin{tikzpicture}[scale=1.5]
                \draw[thick, red] (0.9,-0.7) arc (125:55:-1.56) (-0.9,-0.7) -- (0.9,-0.7);
                \node[red] at (0,-0.85) {$\ket{\psi}$};
                \draw[thick, red] (0.5,-0.7) -- ++(0,0.84) (0.5,0.47) -- (0.5,1.27);
        \node[draw, thick, rectangle,minimum width=0.9cm,minimum height=0.7cm, fill=none] (A) at (-0.5,-0.27) {${M}_{A}$};
        \node[draw, thick, rectangle,minimum width=0.9cm,minimum height=0.7cm, fill=none] (B) at (-0.5,0.87) {${M}_{B}$};
        \draw [thick] (-0.8,1.95) -- (-0.8,1.75) -- (-0.2,1.75) -- ++(0,0.2) arc (0:180:0.3);
                \node (C) at (-0.5,1.95) {$M_C$}; 
        \node[draw, thick, red, rectangle,minimum width=2.5cm,minimum height=0.5cm, fill=none] (U) at (0,0.3) {$W_{1}$};
        \node[draw, thick, red, rectangle,minimum width=2.5cm,minimum height=0.5cm, fill=none] (U2) at (0,1.43) {$W_{2}$};
        \draw[thick, red] (-0.5,-0.7) -- ++(0,0.188);
        \draw[thick, red] (A) -- ++(0,0.395); \draw [thick,red] (B) -- ++(0,0.383);
        \draw [thick, red] (B) -- ++(0,-0.395); \draw [thick,red] (-0.5,1.748) -- ++(0,-0.15);
                \node[] (label) at (0,-1.4) {(b)};
                \node[] (hidden) at (-1.5,0) {};
  \end{tikzpicture}
  \caption{(a) General three-party process matrix $W \in A_I \otimes A_O \otimes B_I \otimes B_O \otimes C_I$. (b) Since, in our scenarios, $W$ is compatible with the causal order $A\prec B \prec C$, we can also represent $W$ as a ``causal network'', which can be implemented as a quantum circuit ($\ket{\psi}$ is a state, $W_1$ and $W_2$ CPTP maps; lines can represent quantum systems of different dimensions.)}
  \label{fig:w3}
\end{figure}

In this paper, we consider three parties, where the $C$'s output space $C_O$ can be disregarded. The process matrix $W \in A_I \otimes A_O \otimes B_I \otimes B_O \otimes C_I$, which encodes the quantum causal model, is defined on the dual space to the tensor products of the maps. Since both the ``common-cause'' and the ``direct-cause'' scenarios are compatible with the causal order $A \prec B \prec C$, we can also represent the process matrix $W$ as a circuit. (see Fig.~\ref{fig:w3}).

For instance, the coherent superposition of common cause and direct cause, defined in \eqref{eq:coherent-pure}, would consist of $\ket{\psi} =  \ket{\phi^+} \otimes (\ket{0}+\ket{1})/\sqrt{2}$, $W_1$ and $W_2$ being control-SWAPs (where the control is the last qubit, initially in the state $(\ket{0}+\ket{1})/\sqrt{2}$).

We now define the projection $\mathcal L_{A\prec B\prec C} (\cdot )$ onto the linear subspace of process matrices compatible with the causal order $A\prec B \prec C$, which can be derived from the conditions given in Ref.~\cite{araujo_witnessing_2015}:
\begin{multline}\label{eq:projector}
\mathcal L_{A\prec B\prec C} (W) := W - _{C_I} W  + _{B_O C_I} W \\ - _{B_I B_O C_I} W + _{A_O B_I B_O C_I} W.
\end{multline}
$W^{A\prec B \prec C}$ is compatible with the causal order $A \prec B \prec C$ if and only if $W^{A\prec B \prec C} = \mathcal L_{A \prec B\prec C} (W^{A\prec B \prec C})$ holds.

The projection onto the subspace of common-cause process matrices $\mathcal L_{\text{cc}} (\cdot)$ is given by composing the projection $\mathcal L_{A \prec B\prec C}$ with the projection onto processes which have no channel from $A_O$ to $B_I$:
\begin{multline}\label{eq:projector-cc}
\mathcal L_\text{cc} (W) := \mathcal L_{A \prec B\prec C}(W) - _{C_I} \mathcal L_{A \prec B\prec C}(W) \\ + _{C_I A_O} \mathcal L_{A \prec B\prec C}(W).
\end{multline}

\section{Dual cones}
\label{app:dualcone}
Given the definition \eqref{eq:process_representation} of the cone $\mathcal W$, we can characterize the \emph{dual cone} $\mathcal W^*$ of all operators whose product with operators in $\mathcal W$ has positive trace. Remember that $\mathcal W$ is the \emph{intersection} of the cone of positive operators $\mathcal P$ with a linear subspace defined by the conditions for causal order: $\mathcal W := \mathcal P \cap \mathcal L_{A\prec B\prec C}$.

The dual of the linear subspace $ \mathcal L_{A\prec B\prec C}^*$ is its orthogonal complement~\cite{nesterov_interior_1987,araujo_witnessing_2015}
\begin{equation}\label{eq:dual-subspace}
\mathcal L_{A \prec B \prec C}^* = \mathcal L_{A \prec B \prec C}^\perp,
\end{equation}
i.e., the space of operators with a support that is orthogonal to the original subspace.

Additionally, the dual of the intersection of two closed convex cones containing the origin is the convex union of their duals~\cite{nesterov_interior_1987,araujo_witnessing_2015}, so that
\begin{equation}\label{eq:dual-cone-implicit}
\mathcal W^*= (\mathcal P \cap \mathcal L_{A\prec B\prec C})^* = \text{conv}(\mathcal P^* \cup \mathcal L_{A \prec B \prec C}^\perp).
\end{equation}
Since the cone of positive operators is self-adjoint ($\mathcal P^* = \mathcal P$), we can combine \eqref{eq:dual-subspace} and \eqref{eq:dual-cone-implicit} into $\mathcal W^* = \text{conv}(\mathcal P \cup \mathcal L_{A \prec B \prec C}^\perp)$. Explicitly, this means that any operator $Q \in \mathcal W^*$ can be decomposed as
\begin{equation}
\label{eq:dualcone}
\begin{gathered}
Q = Q_1 + Q_2 \\
\text{s.t.}~ Q_1 \geq 0, \quad \mathcal L_{A \prec B \prec C} (Q_2) = 0.
\end{gathered}
\end{equation}

\section{Nonclassicality of causality witnesses}
\label{app:s1s2}
We will now explicitly construct the set of nonclassicality of causality witnesses $\mathcal S_\text{SDP}$. 

The semidefinite relaxation of the direct-cause constraint \eqref{eq:w-direct-cause} in terms of positive partial transposition is (using the shorthand introduced in \eqref{def:notation_trace}):
\begin{equation}\label{eq:w-direct-relax}
(_{C_I B_O} W^{\text{dc}})^{\text{T}_{A_I}} \geq 0.
\end{equation}
The dual cone \eqref{eq:s-direct-relax} to the cone of relaxed direct-cause processes defined by the intersection of $\mathcal W$ with the cone defined in \eqref{eq:w-direct-relax} and the dual cone \eqref{eq:s-common-cause} to the cone of common-cause processes defined by the intersection of $\mathcal W$ with the linear subspace \eqref{eq:w-common-cause} can be constructed in the same way as in Appendix~\ref{app:dualcone}.

The set of witnesses positive on all positive partial transpose operators is a \emph{subset} of entanglement witnesses. Every witness belonging to this set satisfies\footnote{We included the term $S_2$ and $S_3$ although they do not make the witnesses ``more powerful'' to detect entanglement. $S_2$ will become relevant when combining the conditions on direct-cause and common-cause processes in Eq.~\eqref{eq:sdp-witness-explicit2}; $S_3$ is included because it could appear in restricted types of witnesses~\cite{araujo_witnessing_2015}.}:
\begin{equation}\label{eq:s-direct-relax}
\begin{gathered}
S^{\text{dc}} = _{C_I B_O} (S_1^{\text{T}_{A_I}}) + S_2  + S_3 \\ \text{s.t.}~S_1, S_2 \geq 0, \quad  \mathcal L_{A\prec B\prec C} (S_3) = 0.
\end{gathered}
\end{equation}
If $\tr [S^{\text{dc}}\, W] < 0$, this implies that $W$ is not a direct-cause process as defined in Eq.~\eqref{eq:w-direct-cause}. Note that since we are only considering a subset of entanglement witnesses, \emph{the converse does not hold}.

We can now turn to the requirement that $S$ is positive on common-cause processes. Since condition~\eqref{eq:w-common-cause} (corresponding to \eqref{eq:projector-cc} together with positivity) defines a convex cone, we can use the techniques of Appendix~\ref{app:dualcone} to construct the dual cone, of which the witness will be an element. This leads us to write $S$ as
\begin{equation}
\label{eq:s-common-cause}
\begin{gathered}
S^{\text{cc}} = S_4 + S_5\\
\text{s.t.}~S_4 \geq 0, \quad \mathcal L_\text{cc} (S_5) = 0,
\end{gathered}
\end{equation}
where the projection onto the common-cause subspace $\mathcal L_\text{cc}$ is defined in Appendix~\ref{app:causallyordered}. $W$ is \emph{not} a common-cause process as defined in~\eqref{eq:w-direct-cause} if and only if there exists an $S^\text{cc}$ such that $\tr [S^{\text{cc}}\, W] < 0$.

Now, combining both conditions, we can construct a set of operators positive on all mixtures of direct-cause and common-cause processes \emph{only in terms of semidefinite constraints}. To test whether an arbitrary $W$ process is of this type, we can run the following semidefinite program (SDP)~\cite{nesterov_interior_1987}:
\begin{equation}
\label{eq:sdp-witness-explicit2}
\begin{gathered}
\min \tr [S\,W]\\
\text{s.t.} \quad S = _{C_{I} B_O}(S_{1}^{\text{T}_{A_{I}}}) + S_{2} + S_3 = S_{4} + S_{5},\\
S_{1} \geq 0, \quad S_{2} \geq 0, \quad S_{4} \geq 0,\\
\mathcal L_{A \prec B \prec C} (S_3) =  \mathcal L_\text{cc}  (S_5) = 0,\\
\quad \id/d_{O} - S \in \mathcal W^{ *}.
\end{gathered}
\end{equation}
The last condition, where $\mathcal W^*$ is the cone dual to $\mathcal W$ (see Appendix~\ref{app:dualcone}), imposes a normalization on $S$. It gives the nonclassicality of causality $\mathcal C (W) = -\tr [S_{\text{opt}} W]$ the operational meaning of ``generalized robustness'', quantifying resistance of the nonclassicality detectable by $\mathcal S_\text{SDP}$ to \emph{worst possible noise}~\cite{steiner_generalized_2003,araujo_witnessing_2015}. This becomes more intuitive from the dual SDP, given by
\label{app:dualsdp}
\begin{equation}
\label{eq:sdp-w-explicit}
\begin{gathered}
\min \tr [\Omega/d_{O}]\\
\text{s.t.} \quad W + \Omega =  W^{\text{cc}} + W^{\text{dc}},\\ 
(_{C_{I} B_O} W^{\text{dc}})^{\text{T}_{A_{I}}} \geq 0, \quad W^{\text{dc}} \in \mathcal W, \\ 
_{C_{I}} W^{\text{cc}} = _{C_I A_O} W^{\text{cc}}, \quad W^{\text{cc}} \in \mathcal W.
\end{gathered}
\end{equation}
The process $\Omega\cdot d_O/\tr[\Omega]$ can be interpreted as worst-case noise with respect to the optimal witness $S_\text{opt}$, resulting from the SDP \eqref{eq:sdp-witness-explicit2}.

\section{Convexity and monotonicity}
\label{sec:measure}
Here we prove that the \emph{nonclassicality of causality} defined as $\mathcal C (W) := -\tr[S_\text{opt} W]$, which results from the SDP \eqref{eq:sdp-witness-explicit2}, satisfies the natural properties of \emph{convexity} and \emph{monotonicity}, following analogous proofs of Ref.~\cite{araujo_witnessing_2015}.

\emph{Convexity} means that $\mathcal C (\sum_i p_i W_i) \le \sum_i p_i \mathcal C(W_i)$, for any $p_i \geq 0, \sum_i p_i = 1$. Take $S_{W_i}$ to be the optimal witness for $W_i$. Any other witness, in particular the optimal witness $S_W$ for $W := \sum_i p_i W_i$ will be less robust to noise with respect to $W_i$:
\begin{equation}
 \tr[S_{W_i}\, W_i] \le \tr [S_W\, W_i].
\end{equation}
Averaging over $i$ we have
\begin{equation}
 - \tr \left[S_{W} \sum_i p_i W_i \right] \le -\sum_i p_i \tr[S_{W_i} W_i],
\end{equation}
which is exactly the statement of convexity for $\mathcal C$.

\emph{Monotonicity} under local operation means that $\mathcal C(W) \geq \mathcal C(\$(W))$, where $\$(\cdot)$ is the composition of $W$ with local operations.

We wish to show that $- \tr \left [S_{\$(W)} \$ (W) \right ] \le - \tr [S_W W]$. By duality, this is equivalent to
\begin{equation}\label{eq:monotonous-dual}
- \tr \left [\$^* \left(S_{\$ (W)} \right) W \right ] \le - \tr [S_W\, W],
\end{equation}
where $\$^* (\cdot)$ is the map dual to $\$ (\cdot)$. Eq.~\eqref{eq:monotonous-dual} is satisfied if $\$^* \left(S_{\$ (W)} \right)$ is a witness, i.e., is positive on all mixtures of direct-cause and common-cause operators ($\tr \left[\$^* \left(S_{\$ (W)} \right) W^\text{mix} \right] \geq 0$), and is normalized appropriately ($1/d_O - \$^* \left(S_{\$ (W)} \right) \in \mathcal W^*$).

The first condition can be seen to hold by applying duality and using the fact that local operations map any mixture of direct-cause and common-cause processes to a mixture of direct-cause and common-cause processes. The second condition is equivalent to 
\begin{equation}
\tr \left [\id/d_O - \$^* \left (S_{\$ (W)} \right) \Omega \right] \geq 0
\end{equation}
for every process matrix $\Omega$. We apply duality and linearity of the trace to find that
\begin{equation}
\tr \left [S_{\$ (W)} \$(\Omega) \right] \leq \tr [\Omega]/d_O.
\end{equation}
This relation holds because $\$(\cdot)$ maps normalized ordered process matrices to normalized ordered process matrices and $\id/d_O - S_{\$(W)} \in \mathcal W^*$ is the normalization condition for the SDP~\eqref{eq:sdp-witness-explicit2}.

The condition of \emph{discrimination} (or \emph{faithfulness}), which would mean that $\mathcal C (W) \geq 0$ \emph{if and only if} the process matrix is not a mixture of direct-cause and common-cause processes \eqref{eq:w-separable}, is \emph{not satisfied}. Since we relied on a relaxation of the direct-cause condition by using the positive partial transpose criterion, there are processes which are not a mixture satisfying \eqref{eq:w-separable} but for which the nonclassicality of causality is zero. 

Therefore, the nonclassicality of causality is not a \emph{faithful measure} of the nonclassicality of the causal structure. This is reasonable, since finding such a measure would be equivalent to finding a fully general \emph{entanglement criterion}---a problem known to be computationally hard~\cite{gurvits_classical_2003}.

\section{Experimental constraints on witnesses}
\label{app:exp-constraints}
In this appendix, we give the explicit form of the experimental constraints mentioned in the main text. When using a constrained class of witnesses, the  value $- \tr [S^{\text{restricted}}_{\text{opt}} W_{\text{coherent}}] $. can be interpreted as the \emph{amount of noise} tolerated before the \emph{constrained set of witnesses} becomes incapable of detecting the nonclassicality of causality of $W_\text{coherent}$.

A simple example of a restriction simplifying the experimental implementation consists in disregarding the space $C_{I}^{(1,2)}$, i.e., to have $S = _{C_{I}^{(1,2)}}S$ as an additional constraint. The nonclassicality of causality is \emph{unaffected} by this restriction, which shows that the input spaces $C_{I}^{(1,2)}$ do not carry any additional information about the nonclassicality of causality.

The constraint for the witness to consist only of non-demolition measurements is:
\begin{multline}
\label{eq:witness-observation}
S^{\text{ndmeas}} = \sum_{ijl} \alpha_{ijl} (\id +\sigma_{i}^{A_{I}})\otimes (\id + \sigma_{i}^{A_{O}})\\
 \otimes (\id +\sigma_{j}^{B_{I}}) \otimes (\id +\sigma_{j}^{B_{O}}) \otimes E_{l}^{C_I},
\end{multline}
where $\sigma_k$ ($k=1,2,3$) are the qubit Pauli matrices and $E_l$, $l=1,\dots, 8$ is an arbitrary basis of projectors on $C_I$'s three qubits.

The constraint for the witness to only consist of unitary operations\footnote{Note that according to definition of Ref.~\cite{ried_quantum_2015}, unitary witnesses should also be considered as ``observations'' although operationally they are standardly understood as interventions.} for $A$ and $B$ is:
\begin{multline}
\label{eq:witness-unitary}
S^{\text{unitary}} = \sum_{ijl} \beta_{ijl} \ket{U_{i}^{ * }}\rangle\langle\bra{U_{i}^{ * }}^{A_{I}A_{O}} \\
\otimes \ket{U_{j}^{ * }}\rangle\langle\bra{U_{j}^{ * }}^{B_{I}B_{O}} \otimes E_{l}^{C_I},
\end{multline} 
where $i,j = 1,\ldots, 10$ indexes a basis\footnote{This is because there are ten linearly independent projectors on CJ-vectors for unitaries acting on qubits~\cite{araujo_witnessing_2015}.} of the CJ-vectors (see Appendix~\ref{app:cj}) of unitaries. 

\begin{table}[htb]
\caption{\label{tab:gr-restricted-witnesses}Constrained nonclassicality of causality for different types of constraints on $S$, in descending order.}
\centering
\begin{tabular}{lc}
\hline
Constraint on the witness $S$ & $\mathcal - \tr[S\,W^\text{coherent}]$\\
\hline
No constraint & 0.2278\\
Discarding $C_I^{(1,2)}$ & 0.2278\\
Unitary operations $A,B$ & 0.1686\\
ND measurement $A,B$ & 0.0732 \\
\hline
\end{tabular}
\end{table}

\section{Definition of direct-cause processes and relationship to the definitions of Ref.~\cite{maclean_quantum-coherent_2016}}
\label{app:def-cause-effect}
Since Ref.~\cite{maclean_quantum-coherent_2016} considers two party case, we can merge $B$ and $C$ to make our scenario comparable to the one of Ref.~\cite{maclean_quantum-coherent_2016}. More precisely, $B_I$ and $C_I$ are relabeled as $B'_I$ and $B_O$ is disregarded, eliminating the necessity to trace over $B_O$ and $C_I$. The condition for direct-cause processes \eqref{eq:w-direct-cause} then becomes
\begin{equation}
\label{eq:w-direct-cause-app}
 W^{\text{dc}} = \sum_i p_i \rho_i^{A_{I}} \otimes \tilde {W}_i^{A_{O} B_{I}'},
\end{equation}
which implies that the states given to $A$ and the channel connecting $A$ and $B$ can be \emph{classically correlated}. 

\begin{figure}[htp]
  \centering
    \begin{minipage}{.28\linewidth}
    \includegraphics{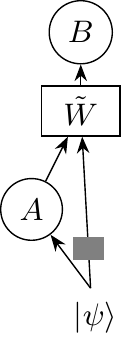}
    \end{minipage}\begin{minipage}{.33\linewidth}
    \includegraphics{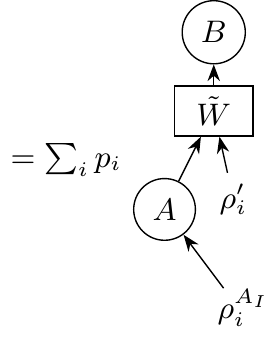}
    \end{minipage}\begin{minipage}{.33\linewidth}
    \includegraphics{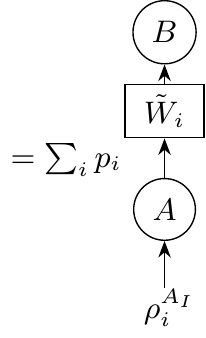}
    \end{minipage}
  \caption{Quantum causal models respecting the extended ``direct-cause'' condition \eqref{eq:w-direct-cause-app} can be thought of as a general channel with \emph{classical} memory (left), or equivalently as a convex combination of direct-cause processes with no memory (right). $\tilde W$ and $\tilde W_i$ are general quantum channels, $\ket{\psi}$ an arbitrary quantum state and the gray square represents a fully dephazing channel (in an arbitrary basis).}
  \label{fig:class-direct-cause}
\end{figure}

In the terminology of DAGs this convex mixture would correspond to tracing over a (hidden) classical\footnote{Strictly speaking, it just needs not to produce any entanglement between $A_I$ and $(A_O, B_I)$, see Fig.~\ref{fig:class-direct-cause}.} common cause between $A$ and $B$. An alternative, more restricted definition would exclude such classical correlations, i.e.,
\begin{equation}
\label{eq:w-direct-cause-alternative-app}
W^{\text{dc}}{} = \rho^{A_{I}} \otimes \tilde {W}^{A_{O} B_{I}'}.
\end{equation}
It is used in Ref.~\cite{maclean_quantum-coherent_2016}. To make the difference apparent, consider the convex mixture of two direct-cause processes between $A$ and $B$ (here, $\dim A_I = \dim A_O = \dim B_I' = 2$):
\begin{multline}
\label{eq:class-with-memory}
W^{\text{mem}} = \frac{1}{4} \ket{0}\bra{0}^{A_I} (\id^{A_O B_I'} + \sigma_z^{A_O} \sigma_z^{B_I'}) \\ 
+  \frac{1}{4} \ket{1}\bra{1}^{A_I} (\id^{A_O B_I'} - \sigma_z^{A_O} \sigma_z^{B_I'}),
\end{multline}
where the tensor products between the Hilbert spaces are implicit. $W^\text{mem}$ classically correlates the channel between $A_O$ and $B_I'$ (a classical channel with or a without bit flip) to the state in $A_I$, as shown in Fig.~\ref{fig:class-direct-cause}. It is of the type \eqref{eq:w-direct-cause-app} but \emph{not} of the type \eqref{eq:w-direct-cause-alternative-app}.

In Ref.~\cite{maclean_quantum-coherent_2016}, \eqref{eq:class-with-memory} is not considered to be a direct-cause process, nor a convex mixture (called ``probabilistic mixture'') of direct-cause and common-cause processes. It is instead termed a ``physical mixture'' of common-cause and direct-cause processes. 

We instead use the broader definition \eqref{eq:w-direct-cause-app} because we ultimately intend to study convex combinations of common-cause and direct-cause processes \eqref{eq:w-separable}, which means we should also allow for convex combinations of direct-cause processes. The restricted definition \eqref{eq:w-direct-cause-alternative-app} for direct-cause processes would lead to consider a convex combination of a direct-cause and a common-cause process to be a ``probabilistic mixture'', but \emph{not} a convex combination of \emph{two cause-effect processes}.

Finally note that the class of processes, which, when post-selected on CP maps being implemented at $B_I'$, result in an entangled conditional process on $A_I A_O$, is defined to be ``coherent mixtures'' in Ref.~\cite{maclean_quantum-coherent_2016}. All of these ``coherent mixtures'' are nonclassical in our terminology (the processes that can be decomposed as \eqref{eq:w-separable} never result in an entangled conditional process on $A_I A_O$). It is not clear whether the converse is true.

\section{Issues in defining a quantum ``observational scheme''}
\label{app:observations}
Ried et al.~\cite{ried_quantum_2015} define the ``observational scheme'' (as opposed to the ``interventionist scheme'') on a quantum causal structure as composed of operations satisfying the ``informational symmetry principle''. We examine the subtleties and issues involved in this definition, in particular regarding the dependence on the initially assigned state.

Ref.~\cite{ried_quantum_2015} assumes that before the observation, one assigns the (epistemic) state $\rho_{A_I}$ to the system coming into $A$'s laboratory. A quantum operation (described by the Choi-Jamio\l{}kowski representation of the quantum instrument~\cite{davies_operational_1970} $\{M_A^{i}\}$, where $i$ labels the outcome) is applied. This updates the information about the outgoing state $\rho_{A_O}^{(i)}$ \emph{but also} (through retrodiction) about the incoming state $\rho_{A_I}^{(i)}$. These states are found by applying the update rules~\cite{leifer_formulation_2013}:
\begin{align}
\rho_{A_O}^{(i)} &= \frac{\tr_{A_I} [M_A^{i} \cdot \rho_{A_I} \otimes \id_{A_O}]^\text{T}}{\tr [M_A^{i} \cdot \rho_{A_I} \otimes \id_{A_O}]},\\
\rho_{A_I}^{(i)} &= \frac{\tr_{A_O} \left[(\sqrt{\rho_{A_I}}\otimes \id_{A_O}) M_A^{i} (\sqrt{\rho_{A_I}}\otimes \id_{A_O}) \right]}{\tr \left[(\sqrt{\rho_{A_I}}\otimes \id_{A_O}) M_A^{i} (\sqrt{\rho_{A_I}}\otimes \id_{A_O}) \right]}.
\end{align}

The informational symmetry principle holds if and only if after the operation, the states assigned to the incoming and outgoing systems are the same:
\begin{equation}\label{eq:informational-symmetry}
\rho_{A_I}^{(i)} = \rho_{A_O}^{(i)}.
\end{equation}
For Ried et al., an instrument for which this informational symmetry holds is \emph{defined} to be an ``observation''~\cite{ried_quantum_2015}. In this sense, there can obviously be ``non-passive'' observations such as non-demolition measurements. Any non-demolition measurement in a basis in which the initially assigned state $\rho_{A_I}$ is \emph{diagonal} will be an observation in this sense. This matches the intuition that a \emph{classical} measurement only \emph{reveals} information and does not disturb the system.

If one wishes to implement measurements in \emph{arbitrary bases}, the \emph{only} initially assigned state which results in informational symmetry is the maximally mixed state $\rho_{A_I} = \id/d$~\cite{ried_quantum_2015}. This shows how problematic the definition of observational scheme is, since it not only crucially depends on an initial (epistemic) assignment $\rho_{A_I}$ but also because there is \emph{only one} such assignment which allows all measurements to be ``observations''---which tolerates no amount and no type of noise. In this sense, as soon as the experimenter \emph{changes her beliefs} about the incoming state \emph{in any way}, she will be intervening on the system, not merely observing it. 

Leaving aside these interpretative difficulties, it is interesting to realize that operations which are \emph{unitary} also turn out to be ``observations'' if the initially assigned state is $\rho_{A_I} = \id/d$: for a unitary operation, $\rho_{A_I}^{(i)} = \rho_{A_O}^{(i)} = \rho_{A_I} = \id/d$. The unitary provides exactly the same information about input and output states, namely \emph{none}.

Finally, note that both the framework of Ref.~\cite{ried_quantum_2015} and the one we developed rely on the assumption that quantum theory is valid and the correct operations were implemented---the analysis is \emph{device-dependent}. This means that any ``quantum advantage'' in inference will not be based on \emph{mere correlations} in the sense of a conditional probability distribution of outputs given inputs. This makes the comparison with the power of classical causal models somewhat problematic.

\bibliographystyle{linksen}
\bibliography{physics-njp}
\end{document}